\documentclass{optica-article}

\journal{oe}
\usepackage{lineno}
\begin{document}

\title{Linear Optical Random Projections Without Holography}

\author{Ruben Ohana,\authormark{1} Daniel Hesslow,\authormark{2,3} Daniel Brunner,\authormark{3} Sylvain Gigan\authormark{4} and Kilian M\"uller\authormark{3,*}}

\address{\authormark{1}Center for Computational Mathematics, Flatiron Institute, 162 Fifth Avenue, New York City, USA\\
\authormark{2}LightOn, 3-5 Impasse Reille, 75014 Paris, France\\
\authormark{3} Institut FEMTO-ST,  Universit\'e de Franche-Comt\'e - CNRS (UMR 6174), 25030 Besan\c{c}on, France.\\
\authormark{4}Laboratoire Kastler Brossel, ENS–Université PSL, CNRS, Sorbonne Université, College de France, 24 Rue Lhomond, F-75005, Paris, France}

\email{\authormark{1}rohana@flatironinstitute.org\\
\authormark{*}kilian@lighton.ai} %% email address is required


\begin{abstract}
We introduce a novel method to perform linear optical random projections without the need for holography. Our method consists of a computationally trivial combination of multiple intensity measurements to mitigate the information loss usually associated with the absolute-square non-linearity imposed by optical intensity measurements. Both experimental and numerical findings demonstrate that the resulting matrix consists of real-valued, independent, and identically distributed (i.i.d.) Gaussian random entries. Our optical setup is simple and robust, as it does not require interference between two beams. We demonstrate the practical applicability of our method by performing dimensionality reduction on high-dimensional data, a common task in randomized numerical linear algebra with relevant applications in machine learning.
\end{abstract}

\section{Introduction}
For most intents and purposes, the propagation of light is linear. The input and output fields of an optical system can therefore be described as vectors $x$ and $y$ that are connected via the transmission matrix $M$ of the system: $y = Mx$. This notion and the ubiquitous use of vector-matrix multiplications in all data processing is one reason for the continuous interest in building optical processors. However, since detectors register the intensity of light, and not its complex amplitude, they inevitably introduce the modulus-square non-linearity and detect $|y|^2$, not $y$. All information about the phase of the complex output field $y$ is therefore lost when optical intensity is the measurement variable. Holographic techniques can be used to recover the phase and the linearity of the operation. However, they come at a cost: indispensable for holographic methods is a reference beam that is made to interfere with the output field, which invokes stringent stability requirements. Further, in the case of phase-shift holography~\cite{yamaguchi1997phase} at least three pictures/measurements must be taken to recover the phase, slowing down the data processing rate. Only one image has to be taken to perform off-axis holography~\cite{cuche1999digital}. Yet, this experimental simplification comes at the price of non-trivial digital post-processing has to be performed, and the maximum output dimension is significantly smaller than the number of pixels on the intensity-recording camera since interference fringes for each output feature need to be resolved.

In this work, we present a method to optically perform a linear vector-matrix multiplication without holography, where the matrix elements are independent and identically distributed random variables drawn from a normal distribution. This operation is surprisingly versatile: distances between vectors are approximately conserved as shown by the Johnson-Lindenstrauss~lemma~\cite{johnson1984extensions}, and the field of Randomized Numerical Linear Algebra (RNLA) is exploiting such operations in a variety of ways to be able to solve linear algebra tasks for very high dimensional data~\cite{drineas2016randnla, martinsson2020randomized}. Consequently, an optical processor that performs random projections can lend itself to all these uses, while keeping the well-known advantages of optical computing: since the computation is executed as the light propagates through the optical system, it is entirely passive, very fast, and highly parallel.

Our method requires as few as $2N+2$ images to perform $N$ linear random projections, does not rely on interference with a reference beam, and only requires computationally trivial post-processing. It therefore allows for a very stable and simple design of a fast optical linear processor. To our knowledge, this method presented in the following is unknown in the community.

\section{Presentation of the method}
We start in the most general setting where $M$ is an arbitrary linear transform, and $x$ is an arbitrary input vector. $x_a$ is a fixed input vector, which in the following we refer to as the \textit{anchor vector}, and we require that $Mx_a$ exclusively comprises non-zero elements.
In the following, the complex conjugate is denoted by an overline $\overline{(\ldots)}$, and $\mathfrak{Re}(\ldots)$ is the real part of a complex value. The absolute value $|\ldots|$, multiplication $(\ldots) \odot (\ldots)$, division $(\ldots)/(\ldots)$ and power $(\ldots)^p$ are element-wise operations. We start with the following identity:
\begin{equation} \label{eq:derivation_1}
|M(x_a - x)|^2 = |M x_a|^2 + |M x|^2 - 2\mathfrak{Re}\left(M x_a \odot \overline{M x}\right)
\end{equation}
Next, we construct a new linear transform $\tilde{M}x = (Mx) \odot \exp(-i \arg(Mx_a))$.
$\tilde{M}$ is defined such that $\tilde{M}x_a$ is real and non-negative.
It follows that for any $x$ we have $|\tilde{M}x| = |Mx|$, $M x_a \odot \overline{M x} = \tilde{M} x_a \odot \overline{\tilde{M} x}$, and $\tilde{M}x_a = |\tilde{M}x_a| = |Mx_a|$. This allows us to separate the real part of the product in Eq.~(\ref{eq:derivation_1}) into the product of two real parts:
\begin{align}
|M(x_a - x)|^2 = |M x_a|^2 + |M x|^2 - 2|M x_a| \odot \mathfrak{Re}\left(\tilde{M} x\right)\\
\Rightarrow Lx \coloneqq  \mathfrak{Re}\left(\tilde{M} x\right) = \frac{|M x_a|^2 + |M x|^2 - |M(x_a - x)|^2}{2|M x_a|} \label{eq:derivation_2}
\end{align}
We note that on the left-hand side of Eq.~(\ref{eq:derivation_2}), we have a linear transformation $L$ of the vector $x$, while the right-hand side only contains quantities that can be obtained from intensity measurements in an optical setup. Of course, this derivation only makes practical sense if one can construct a linear transform that is "useful". In the following, we develop how we employ this procedure to create a random matrix with Gaussian i.i.d. entries.

\begin{figure}[ht]
\centering\includegraphics[width=0.9\textwidth]{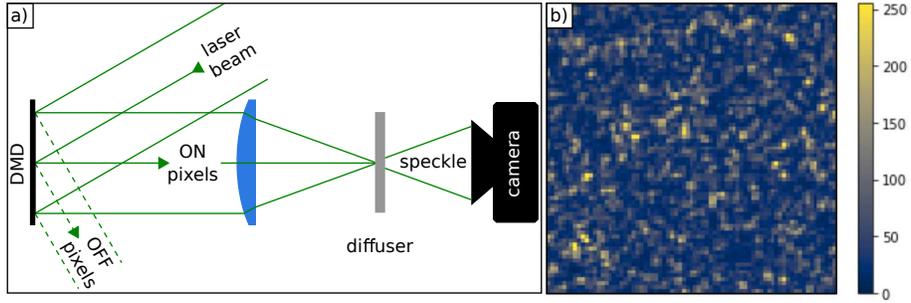}
\caption{Panel a) shows the schematic of the experiment: A collimated laser beam is incident on a DMD. Pixels in the OFF position reflect the light towards a beam dump. Pixels in the ON position direct the light towards a focusing lens. The binary pattern displayed on the DMD is therefore encoded as a 2D binary amplitude pattern in the beam cross-section. In the focal plane of the lens is a thick optical diffuser. Light is multiply and randomly scattered as it propagates to the other side. It is this diffusive medium that results in the random transmission matrix of the overall system. The speckle emanating from the diffuser is captured by a camera. A small section of a speckle is shown in panel b).}
\label{fig:ExpSetup}
\end{figure}

We are using an optical processing unit (OPU) that has been developed by LightOn \cite{brossollet2021lighton}, the principle of which has previously been described in~\cite{saade2016random, ohana2020kernel}. These OPUs have been successfully used in various machine learning settings such as adversarial robustness \cite{cappelli2021ropust,cappelli2022adversarial}, differential privacy \cite{ohana2021photonic}, reservoir computing \cite{rafayelyan2020large} and randomized numerical algebra \cite{hesslow2021photonic}. A schematic drawing of an OPU is shown in Fig.~\ref{fig:ExpSetup}~a). In short, a Digital Micromirror Device (DMD) is used to imprint information (vector~$x$) onto a laser beam. The DMD contains individually controllable mirrors that reflect the light either into a beam block (OFF-state) or towards a diffusive medium (ON-state). We are therefore limited to input vectors with binary entries (0 or 1). As this binary modulated beam propagates through the diffusive medium, it is randomly scattered multiple times. This naturally results in a random optical transmission matrix with normally distributed complex entries~\cite{popoff2010measuring}.

We have found experimentally and in numerical simulations that in the canonical basis, the elements of $L$ as derived in Eq.~(\ref{eq:derivation_2}) have a bias that depends on the choice of the anchor vector~$x_a$. In order to remove this bias and to obtain a matrix whose elements are symmetrically distributed around zero, we work with vectors in the Hadamard basis. This is an orthogonal set of vectors $\{h_i\}$ with binary entries of either -1 or 1. We project them via $L(h_i) = L(h_i^+) - L(h_i^-)$, where $h_i^+$ and $h_i^-$ only have entries 0 or 1 and can therefore be displayed on the DMD. Calculating this difference removes the bias. All of the entries of $h_0$ are equal to 1, making it an obvious choice for the anchor vector $x_a$, since it allows us to display $x_a - x$ on the DMD for all binary vectors $x$, which is necessary for the linear reconstruction.

\section{Experimental Results}
\begin{figure}[ht]
\centering\includegraphics[width=0.5\textwidth]{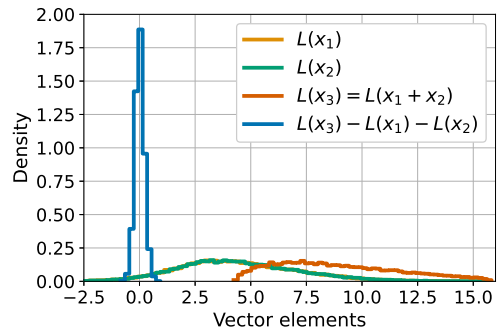}
\caption{Basic test of linearity: The distributions of $L(x_1)$ and $L(x_2)$ are very similar and overlap. The standard deviations are $2.75$ for all $L(x_i)$. The standard deviation of $L(x_3) - L(x_1) - L(x_2)$ is $0.21$.}
\label{fig:LinearityTest}
\end{figure}

We first verify that the transform $L$ defined in Eq.~(\ref{eq:derivation_2}) is indeed linear. For this, we are not obliged to use vectors in the Hadamard basis, but can simply generate three binary input vectors $x_1$, $x_2$, and $x_3$ that fulfill the condition $x_3 = x_1 + x_2$, for which $L(x_3) - L(x_1) - L(x_2)$ should be equal to zero. The result is shown in Fig.~\ref{fig:LinearityTest}. We note that the distributions of $L(x_i)$ are not centered around zero due to the aforementioned bias of $L$ in the canonical basis. The distribution of $L(x_3) - L(x_1) - L(x_2)$, on the other hand, is centered around zero, and its width is about 13~times smaller than those of $L(x_i)$. The remaining finite width is likely due to camera noise.

We then investigate the distribution of the matrix elements of $L$, measured in the Hadamard basis. To obtain the best results, we implement procedures that compensate for imperfections in the experimental setup: Initially, a Gaussian-shaped laser beam illuminates the DMD, resulting in higher light intensity at the center than at the edges. To compensate for this, we assign a random distribution of DMD pixels to each input vector element, creating distributed macro-pixels. Consequently, each macro-pixel receives, on average, equal light intensity. Similarly, we only utilize camera pixels with roughly equal average illumination, as the incident intensity on the detector is not uniform.\\
Next, we account for the size of an average speckle grain on the camera, which has a standard deviation of approximately 0.9~camera pixels. This implies that measured intensities on neighboring camera pixels are correlated, and we use only every third pixel to remove these correlations from $L$.
%As each speckle grain spans more than one pixel, we use every third camera pixel to avoid correlations.
We have measured the correlations between neighboring DMD pixels and confirmed they are negligible.\\
Therefore our system has the flexibility to trade between maximum "cleanliness" -- i.e.  Gaussian distributed and i.i.d. elements -- of the random matrix, or maximum input and output dimensions. Here we choose the former and obtain a system with a maximum input dimension of $\sim 10^5$ and a maximum output dimension of $\sim 3 \times 10^4$.

\begin{figure}[ht]
\centering\includegraphics[width=\textwidth]{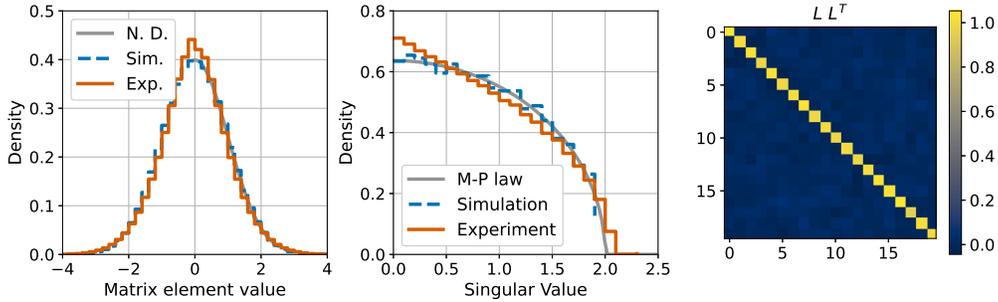}
\caption{\textbf{Left:} A normal distribution (N. D.), simulated data, and the distribution of the elements of the experimental matrix. All distributions are normalized such that their standard deviations are equal to 1. \textbf{Center:} The Marchenko-Pastur (M-P) law for square random i.i.d. matrices -- i.e. the quarter circle law --, compared to the SVDs of simulated data and of the experimental matrix. \textbf{Right:} A $20 \times 20$ subset of $L L^T$, showing that the rows and columns of the experimental transform $L$ are approximately orthogonal. The diagonal and the off-diagonal values of the entire $4096 \times 4096$ matrix are $0.999 \pm 0.023$ and $0 \pm 0.016$, respectively.}
\label{fig:RandMatrixEval}
\end{figure}

Fig.~\ref{fig:RandMatrixEval} shows the results for a $4096 \times 4096$ experimentally measured matrix. We compare these experimental results to a numerical simulation of our method, for which we create $M$ with Gaussian i.i.d. elements. We find that the distribution of the matrix elements of $L$ is Gaussian in both cases, and we attribute the small deviation of the experimental data to remaining and unidentified artifacts of our setup. Furthermore, as described earlier we eliminate correlations of the matrix elements of $L$ by excluding neighboring camera pixels. To verify this method we compare the SVD of $L$ to the Marchenko-Pastur law~\cite{popoff2010measuring, marchenko1967distribution}. In our experience, this is a sensitive probe, with SVDs deviating quickly from the theoretical prediction even for residual correlations. Finally, we show that $L$ is approximately orthogonal by calculating $L L^T$.

\begin{figure}[ht]
\centering\includegraphics[width=0.8\textwidth]{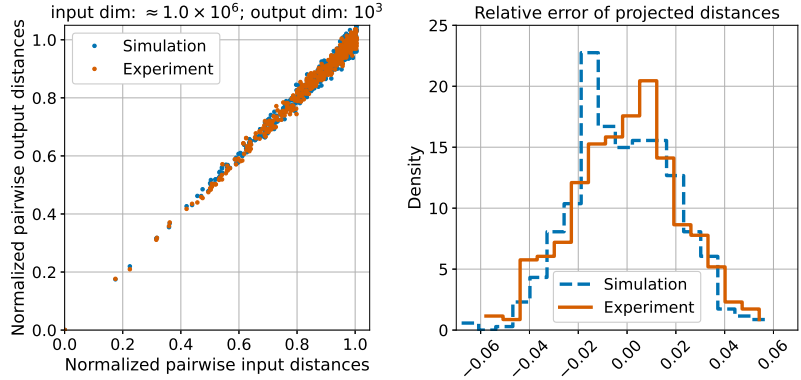}
\caption{Dimensionality reduction via the Johnson-Lindenstrauss lemma. 100 input vectors are projected from the maximal input dimension of our system ($\approx 1.0 \times 10^6$) to a 1000-dimensional space. We have done this numerically and on the experimental setup. In the former, the random matrix has normally distributed i.i.d. elements. \textbf{Left:} The scatter-plot shows that pairwise distances are conserved. \textbf{Right:} A histogram showing the relative errors induced by the compression. Our optical setup performs as well as the numerical projection: In this data set the standard deviation of the simulation is 0.21, and that of the experiment is 0.22.}
\label{fig:JohnsonLindenstrauss}
\end{figure}

In Fig.~\ref{fig:RandMatrixEval} we established that we can obtain a very clean random matrix. To demonstrate that we can also scale to high dimensions, we perform a dimensionality reduction using the Johnson-Lindenstrauss lemma~\cite{johnson1984extensions} using the maximum available input dimension of our system: $912 \times 1140 \approx 1.0 \times 10^6$, equal to the total number of DMD pixels. Consequently, the transform $L$ is now influenced by the inhomogeneous illumination of the DMD, affecting the magnitude of its matrix elements.
We generate 100 input vectors and project them down to a dimension of 1000. The Johnson-Lindenstrauss lemma guarantees that the pairwise distances between vectors are approximately preserved: $\parallel x_i - x_j \parallel \approx \lambda \parallel L(x_i) - L(x_j) \parallel$. For our test, we have normalized $L$ such that $\lambda = 1$. Fig.~\ref{fig:JohnsonLindenstrauss} demonstrates that our experimental setup can carry out the dimensionality reduction without significant performance degradation when compared to a numerical random projection.

So far, we have used input vectors in the Hadamard basis in order to remove the bias of the transform $L$. This makes intuitive sense since the projection of each vector $h_i$ involves the difference of two projections, and any constant bias is thus removed by this subtraction. However, this implies that $4N+1$ images need to be taken for $N$ random projections. Additionally, we have developed a different way to obtain similar results that requires only $2N+2$ measurements, thus speeding up the process: We first note that any affine transformation $\alpha(x)$ can be written as a linear transform plus a constant offset, $\alpha(x) = Fx + C$, and that two linear transforms applied in series result in another linear transform. The idea is then to prepare the input data with an affine transform before applying $L$: $x\rightarrow L(\alpha(x))$. The total bias $C = L(\alpha(0))$ can then be subtracted at the end. Constrained by the binary nature of our DMD, we choose the XOR operation with a constant vector $A$ as our affine operation: $\alpha(x) = x \oplus A$. With this modification our method then is ($\lnot$ denotes the element-wise binary NOT operator):
\begin{align}
    K(x) &\coloneqq L(A \oplus x) - L(0 \oplus x)\\
    &= \frac{|M(x \oplus \lnot A)|^2 - |M(x \oplus A)|^2 + |M(A)|^2 - |M( \lnot A)|^2}{2|M(x_a)|} \label{eq:XORtrick}
\end{align}
We see in the numerator of Eq.~(\ref{eq:XORtrick}) the differences of two pairs of corresponding terms. Following the same logic as above, any constant bias is therefore removed. We have repeated all the tests in this paper using $K$ and obtained similar results.

\section{Discussion}
We have presented a novel method that allows to perform linear random projections without any holographic methods. This simplifies the experimental setup and increases its stability since interference with a reference beam is no longer necessary. In order to obtain the cleanest random matrix, we were restricted to using a subset of the pixels of our camera and need to create macro-pixels on the DMD. Consequently, the size of the accessible random matrix is reduced. However, we have shown with a demonstration of the Johnson-Lindenstrauss lemma that for some RNLA algorithms, a "perfect" random matrix is not necessary, and the system becomes usable at its maximum dimensions. We also want to note that the optical setup could be improved to increase the accessible dimensionality while retaining nice mathematical conditions of the random matrix: output correlations can be minimized, and input and output illuminations be made more homogeneous. Since input SLMs and cameras with $\sim 10^7$ pixels are readily available, random projections with up to a matrix size of $10^7 \times 10^7$ are theoretically possible. Such a matrix would take up on the order of 100~TB in single precision when implemented numerically, showing the enormous potential of using optics in such very high dimensional setting.\\
Implementing our method with any spatial light modulator that acts on the amplitude of light is straightforward, and we hope that it can therefore find immediate application in laboratories that utilize similar experimental configurations. Although we have only developed this method for random projections, we would like to note that it may also find applications for other linear optical transforms. Consider, for example, the optical Fourier transform of a point source centered on the optical axis. The result is an output field with a constant phase. Using the same formalism and this point source as the anchor vector~$x_a$ it follows that up to a constant phase factor $\tilde{M} = M$, and $L$ is the real part of the optical Fourier transform. In general, there may be systems where the liberty in choosing the anchor vector $x_a$, in combination maybe with a gray-scale amplitude SLMs, or SLMs that can control amplitude and phase, could open up further applications.

% \subsection{Supplementary materials in Optica Publishing Group journals}
% Our journals allow authors to include supplementary materials as integral parts of a manuscript. Such materials are subject to peer-review procedures along with the rest of the paper and should be uploaded and described using our Prism manuscript system. Please refer to the \href{https://opg.optica.org/submit/style/supplementary_materials.cfm}{Author Guidelines for Supplementary Materials in Optica Publishing Group Journals} for more detailed instructions on labeling supplementary materials and your manuscript.

% \textbf{Authors may also include Supplemental Documents} (PDF documents with expanded descriptions or methods) with the primary manuscript. At this time, supplemental PDF files are not accepted for partner titles, JOCN and \emph{Photonics Research}. To reference the supplementary document, the statement ``See Supplement 1 for supporting content.'' should appear at the bottom of the manuscript (above the References heading). 

% \section{Backmatter}

% Backmatter sections should be listed in the order Funding/Acknowledgment/Disclosures/Data Availability Statement/Supplemental Document section. An example of backmatter with each of these sections included is shown below.

\begin{backmatter}
\bmsection{Funding}
% Content in the funding section will be generated entirely from details submitted to Prism. Authors may add placeholder text in the manuscript to assess length, but any text added to this section in the manuscript will be replaced during production and will display official funder names along with any grant numbers provided. If additional details about a funder are required, they may be added to the Acknowledgments, even if this duplicates information in the funding section. See the example below in Acknowledgements.
H2020 Future and Emerging Technologies (899794).

\bmsection{Acknowledgments}
% The section title should not follow the numbering scheme of the body of the paper. Additional information crediting individuals who contributed to the work being reported, clarifying who received funding from a particular source, or other information that does not fit the criteria for the funding block may also be included; for example, ``K. Flockhart thanks the National Science Foundation for help identifying collaborators for this work.'' 
We acknowledge support from EU Horizon 2020 FET-OPEN OPTOLogic. S.G. acknowledges funding from the European Research Council ERC Consolidator Grant (SMARTIES-724473).

\bmsection{Disclosures}
% Disclosures should be listed in a separate nonnumbered section at the end of the manuscript. List the Disclosures codes identified on the \href{https://opg.optica.org/submit/review/conflicts-interest-policy.cfm}{Conflict of Interest policy page}, as shown in the examples below:
% \medskip
% \noindent ABC: 123 Corporation (I,E,P), DEF: 456 Corporation (R,S). GHI: 789 Corporation (C).
% \medskip
% \noindent If there are no disclosures, then list ``The authors declare no conflicts of interest.''
The authors declare no conflicts of interest.

% \bmsection{Data Availability Statement}
% A Data Availability Statement (DAS) will be required for all submissions beginning 1 March 2021. The DAS should be an unnumbered separate section titled ``Data Availability'' that
% immediately follows the Disclosures section. See the \href{https://opg.optica.org/submit/review/data-availability-policy.cfm}{Data Availability Statement policy page} for more information.

% Optica has identified four common (sometimes overlapping) situations that authors should use as guidance. These are provided as minimal models, and authors should feel free to
% include any additional details that may be relevant.

% \begin{enumerate}
% \item When datasets are included as integral supplementary material in the paper, they must be declared (e.g., as "Dataset 1" following our current supplementary materials policy) and cited in the DAS, and should appear in the references.

% \bmsection{Data availability} Data underlying the results presented in this paper are available in Dataset 1, Ref. [3].

% \bigskip

% \item When datasets are cited but not submitted as integral supplementary material, they must be cited in the DAS and should appear in the references.

% \bmsection{Data availability} Data underlying the results presented in this paper are available in Ref. [3].

% \bigskip

% \item If the data generated or analyzed as part of the research are not publicly available, that should be stated. Authors are encouraged to explain why (e.g.~the data may be restricted for privacy reasons), and how the data might be obtained or accessed in the future.

\bmsection{Data availability} Data underlying the results presented in this paper are not publicly available at this time but may be obtained from the authors upon reasonable request.

% \bigskip

% \item If no data were generated or analyzed in the presented research, that should be stated.

% \bmsection{Data availability} No data were generated or analyzed in the presented research.
% \end{enumerate}

% \bmsection{Supplemental document}
% See Supplement 1 for supporting content. 

\end{backmatter}

\bibliography{bibliography}

\end{document}